\documentclass[pra, reprint, letterpaper, superscriptaddress]{revtex4-1}
\usepackage{graphicx}
\usepackage{xspace}
\newcommand{\micron}{$\mathrm{\mu}$m\xspace}
\begin{document}

\title{Assembling a ring-shaped crystal in a microfabricated surface ion trap}

\author{Boyan Tabakov}
\affiliation{Sandia National Laboratories, P.O. Box 5800 Albuquerque, NM 87185-1082}
\affiliation{Center for Quantum Information and Control, University of New Mexico, MSC 07–4220, Albuquerque, NM 87131-0001}
\author{Francisco Benito}
\author{Matthew Blain}
\author{Craig R. Clark}
\author{Susan Clark}
\author{Raymond A. Haltli}
\author{Peter Maunz}
\author{Jonathan D. Sterk}
\author{Chris Tigges}
\affiliation{Sandia National Laboratories, P.O. Box 5800 Albuquerque, NM 87185-1082}
\author{Daniel Stick}
\affiliation{Sandia National Laboratories, P.O. Box 5800 Albuquerque, NM 87185-1082}
\affiliation{Center for Quantum Information and Control, University of New Mexico, MSC 07–4220, Albuquerque, NM 87131-0001}

\begin{abstract}
We report on experiments with a microfabricated surface trap designed for trapping a chain of ions in a ring. Uniform ion separation over most of the ring is achieved with a rotationally symmetric design and by measuring and suppressing undesired electric fields.  After minimizing these fields the ions are confined primarily by an rf trapping pseudo-potential and their mutual Coulomb repulsion. The ring-shaped crystal consists of approximately 400 Ca$^+$ ions with an estimated average separation of 9 \micron.
\end{abstract}

\maketitle

\section{Introduction}
The primary motivation for developing segmented surface ion traps is to create scalable structures that support the implementation of quantum algorithms.  By moving the trapped ions in a prescribed routine and applying appropriate gates, these structures form a quantum charge-coupled device (QCCD) \cite{kielpinski} capable of universal quantum computing. The fabrication techniques used to realize these surface traps also enable other geometries with desirable properties. In the case reported here, a trap with a nearly circularly symmetric rf pseudo-potential is designed to hold a ring-shaped chain of equidistant ions.  Ring geometries have been previously demonstrated for studying electron capture \cite{Bliek96}, storing protons and helium ions \cite{Church69}, and studying phase transitions \cite{Waki92}.  In these cases the rings were large ($>$1 cm diameter), employed a 4-rod geometry, and did not have axial control electrodes, but relied on the rf pseudo-potential and Coulomb repulsion to hold the ions in place.  The objective of our experiments was similarly to trap a ring of ions, but instead use a micro-fabricated surface trap with axial control electrodes to minimize electric field deviations which disrupt the circular symmetry of the ion positions. The causes for such deviations include loading holes, gaps in control electrodes, and surface contaminants. These deviations cannot be entirely eliminated (for example, control electrode gaps cannot be made infinitely thin), but they may be minimized in design and by applying appropriate voltages to the control segments. While a ring geometry is not topologically similar to the QCCD array because it lacks features necessary to shuttle ions in a 2-dimensional lattice, it enables experiments from diverse fields. These include simulations of Ising models with long chains \cite{ising_spins}, Hawking radiation \cite{hawking_radiation}, and studies of quantum phase transitions \cite{phase_transitions}, solitons \cite{solitons}, and time crystals \cite{time_crystals}. 

\section{Measuring stray fields}
Undesired electric fields on an ion trap can result from various sources, including laser charging on the trap surface, contaminants, imperfections in fabrication, and uncontrolled charges elsewhere in the vacuum chamber. These fields can vary in magnitude at the level of several thousand V/m \cite{doepaper} and can be reduced by applying compensating voltages to the control electrodes.  A procedure to reduce the fields at N locations would first require simulations of the field throughout the ring produced by a voltage applied to each trap electrode. Then the actual field would have to be measured in three orthogonal directions at each location. Finally, the 3N linear equations which eliminate the electric fields would be solved and the corrective voltages applied to a minimum of 3N electrodes. The experiment described here focuses on correcting only the axial electric fields which are tangential to the ring, at a subset of 39 equidistant positions.  We find that this spatial resolution (90 \micron) is adequate for reducing the axial fields and achieving a near-uniform ion spacing, which is expected given the ion height of 82 \micron.

\section{Trap layout}
\begin{figure}[h]
 \centering
\includegraphics[scale=0.6,bb=0 0 395 394]{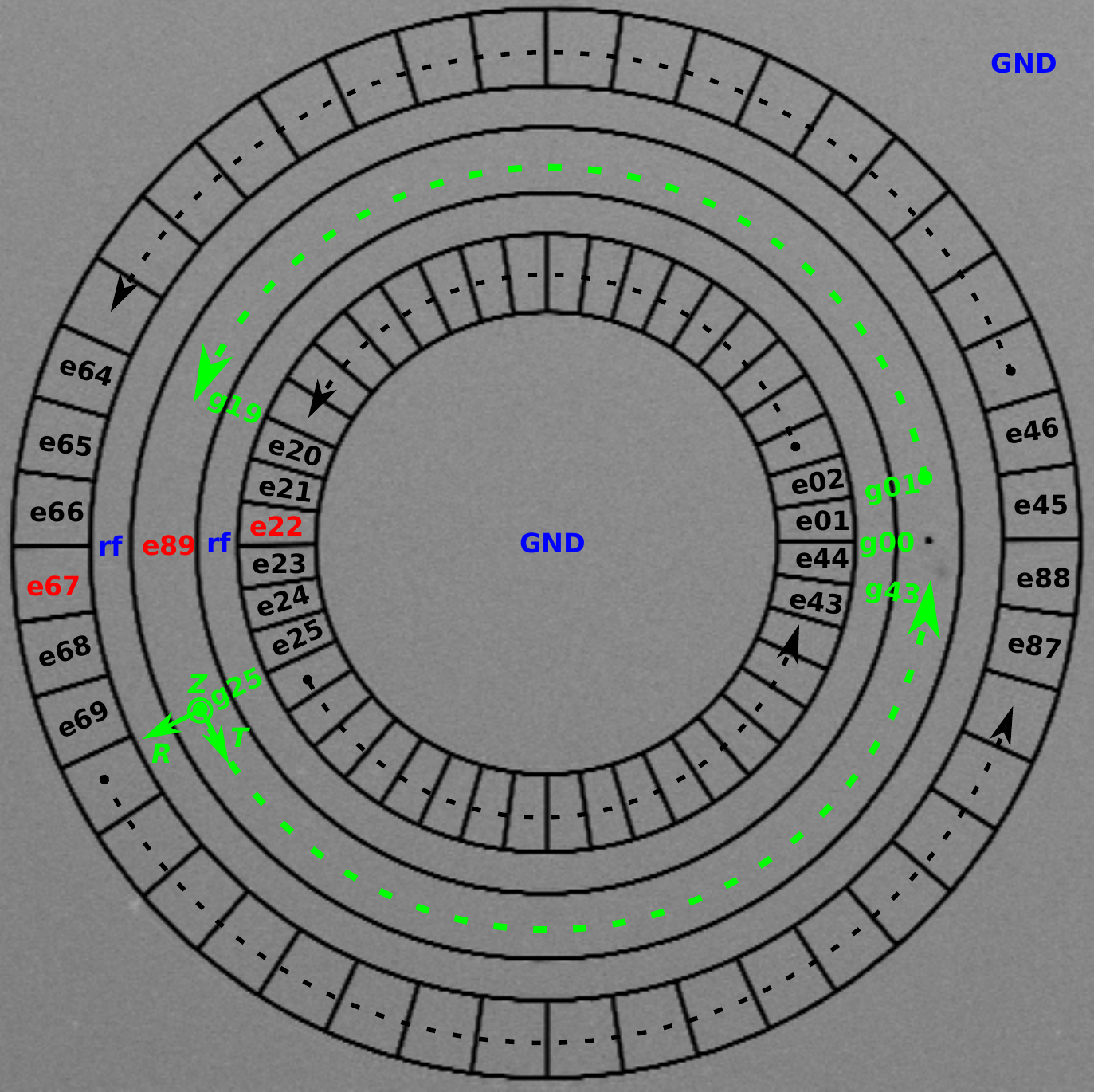}
% ring_layout_v2.pdf: 395x394 pixel, 72dpi, 13.93x13.90 cm, bb=0 0 395 394
\caption{Device layout and labeling. Of the 89 available control electrodes, 86 perform as expected (black labels and dashed arcs), and  $e22$, $e67$, and $e89$ (red labels) are shorted. The trapping volume is above the central control electrode, $e89$, and is designated by the green dashed arcs. The irregularities (see Sec. IV) at the loading hole and the diametrically opposite location result from reduced control over the ion motion at those locations. A trapping site on a radially oriented gap bisector's intersection with the trapping volume minimum is referred to as a ``$g$'' site, with the loading hole at $g00$. An $R-T-Z$ reference frame is shown at $g25$.}
\label{fig:layout}
\end{figure}

A toroidal trapping volume is attained with the electrode layout shown in Fig. \ref{fig:layout}.

At the ring center is a circular ground plane, neighbored by 44 inner control electrodes.  Just outside are two RF electrodes on either side of a central control electrode below the trapping volume.  Next there is an outer set of 44 control electrodes surrounded by an outer ground plane. The rf lead extends outward from the rf ring electrodes on a buried metal level, and at a position 1.7 mm from the trap center it is routed to the top metal level to reduce capacitance.  On the top layer, the gap width between all electrodes is 7 \micron. The trap is designed for back-loading through a 10 \micron  diameter hole, centered at the intersection of a circle of radius 625 \micron and the radial line bisecting the gap between two pairs of inner and outer control electrodes.

The circular symmetry of the trap makes it natural to work in a reference frame centered at a trapped ion (Fig. \ref{fig:layout}). In such a frame, the tangential ($T$) direction (positive counter-clockwise) is the analogue of the axial direction for a linear trap, with the caveat that $\mathbf{\hat{T}}$ is different for each trapping location. Likewise, the radial ($R$) direction (radially outward from the circle center) is different for each location, while the direction perpendicular to the trap plane ($Z$) is universal. The latter two are the radial directions for a linear trap. 

\section{Stray field correction} 
Simulations show that the gaps between control electrodes create only minimal perturbations to the rf pseudo-potential, which are dominated by the Coulomb repulsion between neighboring ions.  Under these conditions, the distance between ions in the ring should be equal. Imaging an actual chain when control electrodes are all set to the same voltage (0 V), however, shows that the ion spacing is irregular.  Ions are bunched in several locations and absent from others, as determined by stray electric fields that break the circular symmetry.

For an arbitrary trapping location, the tangential component of the stray field is measured by scaling the control voltages and measuring the shift of the ion position. For the tangential secular frequency $\omega_T$ of a single trapped ion, $\omega_T^2\propto\alpha$ holds, where $\alpha$ is a scaling factor for a particular set of applied control voltages. The tangential component of the ion potential is $\phi_T = \frac m  {2q}  \omega_T^2 x^2-E_T x$,  where $m$ is the ion mass, $E_T$ is the stray field tangential component, $q$ is the ion charge, and $x$ is the ion's tangential displacement. At equilibrium, the displacement is a function of the stray field $E_T$ : $x =\frac{ E_T q} {m \omega_T^2}\frac{1}{\alpha}$. Displacement measurements for several scalings of the trapping potential along with a single secular frequency measurement can be fitted to yield the tangential electric field value. To assess the field components in the other two orthogonal directions, we apply corrective fields (simulated to be homogeneous in radial and vertical direction, respectively) in the $R$ and $Z$ directions to minimize the secular motion peaks seen with a parametric scan \cite{shankara}, \cite{parametric_resonance}, and also minimize the correlation signal in a time of arrival experiment \cite{berkeland}.

%For all trapping locations around the ring, the stray field can be measured, and one or more electrodes can be assigned to produce a correcting field at that location. Any such correction, however, will alter the trapping potential everywhere else in the trapping volume. This is taken into account by computing the effect of all participating electrodes at all correction sites simultaneously. We assume that that changes in the field are smooth enough so that we correct in between trapping sites for which we calculate the correction. Since there are 44 pairs of electrodes plus a center control electrode, there are, in general, at most 89 corrections that can be applied to the field simultaneously. These corrections can be chosen judiciously: stray field can be suppressed in three directions at 29 trapping locations, or the field in a specific direction can be suppressed at 89 points, given that all participating electrodes generate a field that has a component in the direction of interest(a counter-example: the central control electrode can not be used to provide field in the tangential direction at any trapping location). %

\begin{figure}[h]
 \centering
 \includegraphics[scale=.45,bb=0 0 576 432]{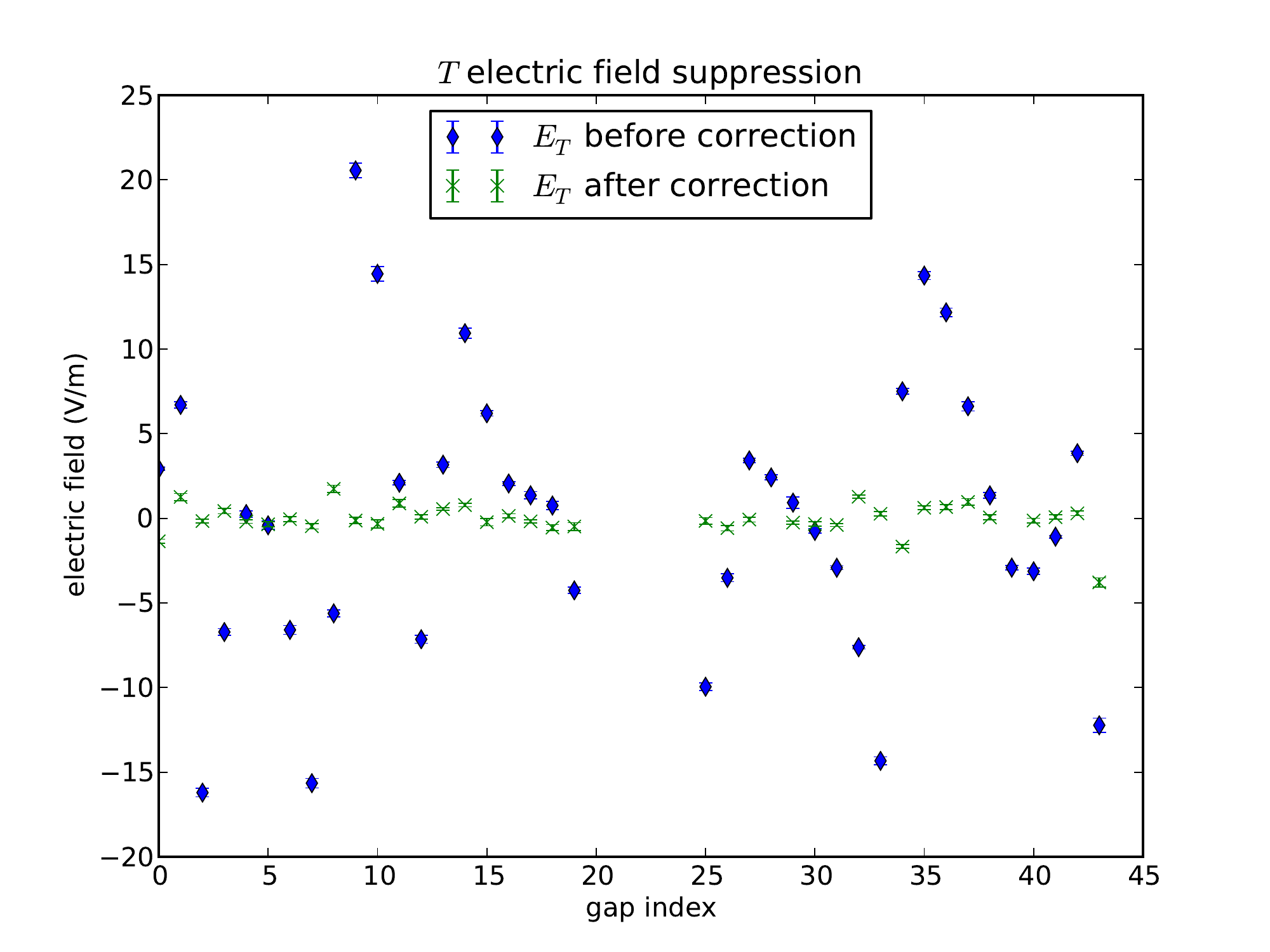}
 \caption{Tangential field suppression. The error bars (small on this scale to be readily visible) are propagated from secular frequency and ion position uncertainties. The field at $g28$ and $g39$ is not calculated after correction due to difficulty in measuring the secular frequency.}
 \label{fig:tf}
\end{figure}

To achieve an equidistant chain, the tangential field is measured at 39 locations. The field that a single electrode produces extends to other locations, therefore 39 equations have to be solved simultaneously so that the total generated field (according to the model) cancels the measured field at each location. We apply the outlined measurement to ions trapped at locations $g00-g19$ and $g25-g43$ (see Fig. \ref{fig:layout}), inclusive. We exclude locations $g20-g24$ due to shorted electrodes and therefore an inability to correct the axial electric field. Measurements of the tangential field components, before and after correction (Fig. \ref{fig:tf}), show that the simultaneously applied correction contributes to homogenizing the tangential field at all measurement locations.

\begin{figure}[h]
 \centering
 \includegraphics[scale=.14]{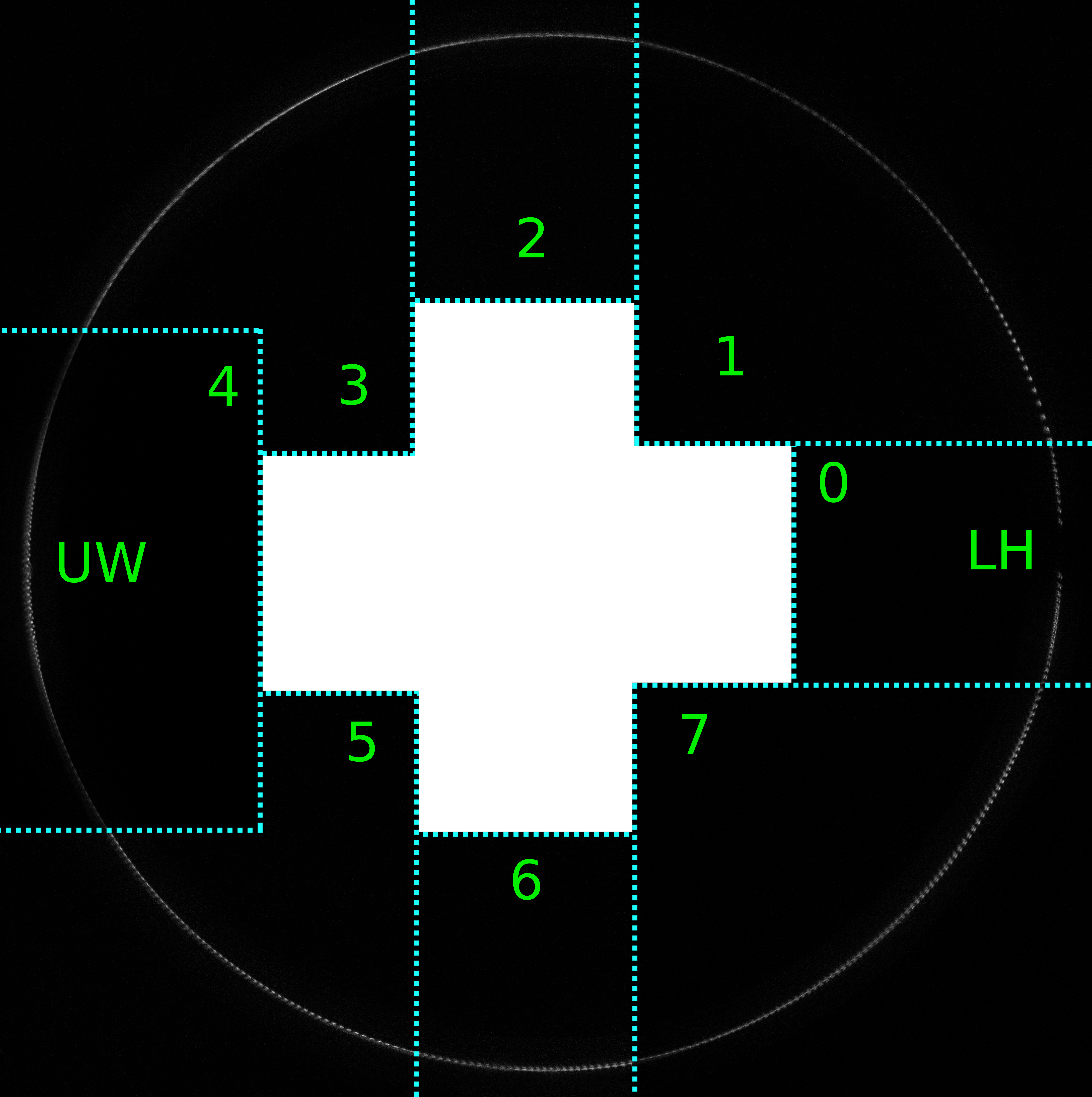}
 \caption{A composite image of a full ring after correction.  The stitched image comprises separate images of eight octants (labeled by the numbers shown). Without correcting the background fields, ions would primarily fall in an unwanted potential well (marked ``UW'') in proximity to the shorted electrodes $e22$ and $e67$ (see Fig. \ref{fig:layout}). With the correction, ions occupy the trapping volume except for above the loading hole (marked ``LH''). Uneven illumination in octants 1, 2, 3, 5, 6, and 7 is due to the gaussian profile of the cooling and re-pump beams. In octant 4, uneven illumination is primarily a consequence of having limited control over electrodes $e22$ and $e67$, thus preventing field measurement and subsequent correction. The chain interruption at the loading hole is due to deficiencies in the model used to calculate the correction.}
 \label{fig:full}
\end{figure}

A demonstration of the effect of tangential field homogenization is the long chain (Fig. \ref{fig:full}) 

\begin{figure}[h]
 \centering
 \includegraphics[scale=.45,bb=0 0 576 432]{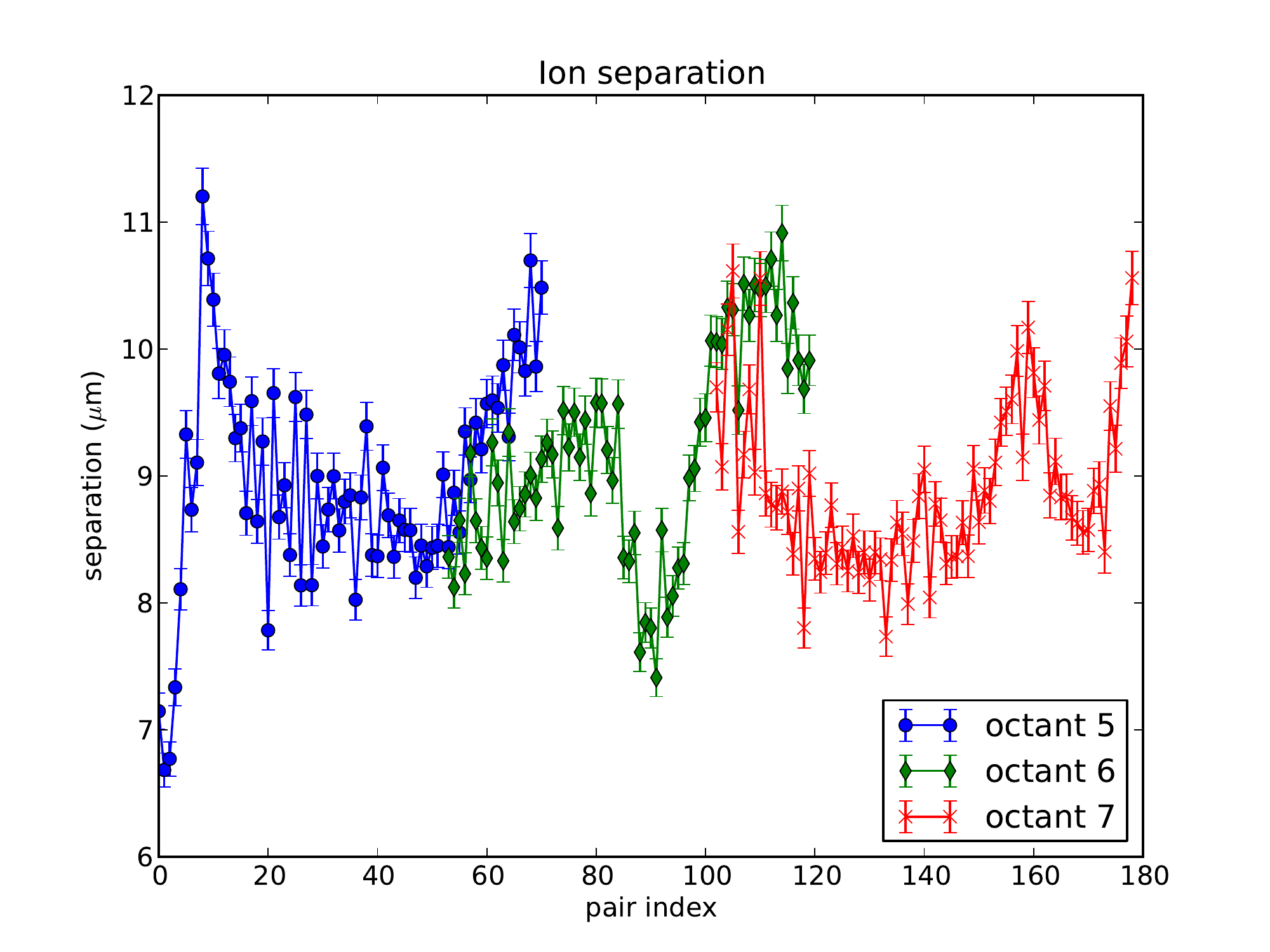}
 \caption{Ion separation in three neighboring octants for which all ions are resolved. Overlap is inferred from the composite image of the full ring (Fig. \ref{fig:full}). Imaging system and cooling beam imperfections prevent resolving all ions in the remaining five octants.  The error bars are based on the uncertainty of the total magnification, but the positions are not corrected for aberations at the edge of the collection optic.}
 \label{fig:separation}
\end{figure}

obtainable only after applying the correction. The magnification of the imaging system balances the competing requirements of individual ion resolution while capturing a large area. An image with the ring center in a corner covers almost a quarter of the trapping volume. To image the whole chain, overlapping images are taken in each octant as the camera is moved in 45$^\circ$ increments along the circle. The obtained images are composed manually because the total ion count changes slightly during this process.  To numerically assess the tangential field homogeneity after stray field suppression at intermediate locations between the electrode gaps, we analyze images in which all ions are resolved. The ion spacing is plotted (Fig. \ref{fig:separation}) for octants 5, 6, and 7.

In the rest of the octants not all ions could be resolved due to a combination of cooling and imaging imperfections.

\begin{figure}[h]
 \centering
 \includegraphics[scale=.45,bb=0 0 576 432]{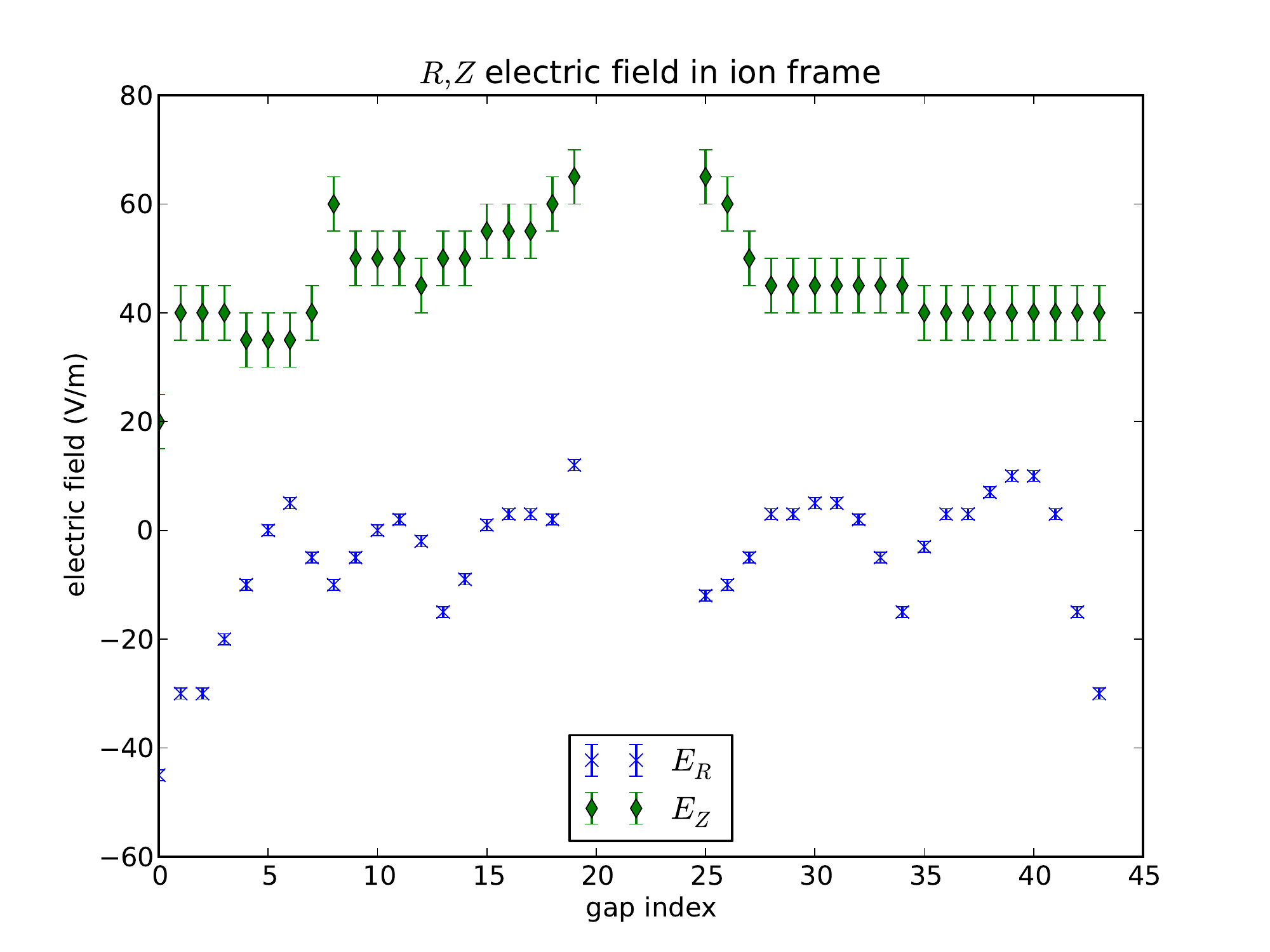}
 \caption{Radial field measurements at 39 locations, without applying tangential corrections. No attempt was made to counter the stray field in these directions, due to restricted degrees of freedom (too few electrodes). The size of the error bars correspond to the change in applied field that did not produce an observable effect on the secular motion peaks or time of arrival amplitude and phase, corresponding to 1 V/m in $\mathbf{\hat{R}}$ and 5 V/m in $\mathbf{\hat{Z}}$.}
 \label{fig:rf}
\end{figure}

To better appreciate stray fields details, the radial electric field components ($R$ and $Z$) are measured, as shown in Fig. \ref{fig:rf}. The error bars in the measured stray field are chosen to be the largest field applied in the respective direction that did not produce an appreciable change in the minimized feature being monitored (secular motion peaks amplitude or time of arrival fit amplitude and phase). These data did not enter in the calculation of the tangential correction that produced the long chain.

\section{Trap simulation and performance}
The modeled pseudo-potential minimum is a circle of radius 624 \micron at 82 \micron above the trap surface. The model is created by meshing the surface geometry with CUBIT \footnote{https://cubit.sandia.gov/}, feeding the mesh to CPO \footnote{http://simion.com/cpo/} for electric field estimation, and manipulating the calculated field with Mathematica \footnote{http://www.wolfram.com/mathematica/} for parameter extraction. The trap geometry is optimized to increase trap depth and minimize the impact of the rf lead breaking the circular symmetry. The diameter of the ring is chosen to trap a circular crystal of hundreds of ions spaced at tens of \micron. For 400 ions spaced at 10 \micron, the generated trapping strength is 360 kHz for each ion. A 10 \micron diameter loading hole is chosen because it only deforms the pseudo-potential with a 9 kHz local trapping potential, significantly smaller than the ion-ion trapping strength. Due to limited computing resources, this consideration is not reflected in the large scale simulation used to obtain the correction, resulting in the observed gap in the chain at the loading hole after correction.
%The effects of the loading hole and gap were not captured by the model. 

Away from the loading hole, the secular frequency in the tangential direction ($\omega_T$) predicted by the simulations agreed with measurement to within a few percent. In most single ion experiments, control voltages are used to rotate the radial principal axes $15^\circ$ and achieve an axial trapping frequency of $\omega_T=2\pi\times 290$ kHz. From measured values of $\omega_R=2\pi\times 2.12$ MHz and $\omega_Z=2\pi\times 2.17$ MHz, by comparing to the model, the estimated voltage delivered to the trap rf electrodes at frequency $\Omega = 2\pi\times52.9$ MHz is 80 V, and the estimated trap depth for Ca$^+$ is $17$ meV.  A heating rate of 1.06(24) quanta/ms is measured by using sideband ratio techniques \cite{turchette}, for $\omega_T\approx2\pi\times 0.9$ MHz.

\section{Trap fabrication and experimental setup}
Surface ion traps designed to accommodate multiple ions in unique spatial arrangements, such as a ring, typically require the use of multiple, segmented trap electrodes. To realize these arrangements and provide independent spatial control of many ions, multiple layers of routing metal are necessary. While two metal level surface traps have been demonstrated \cite{y_trap}, \cite{gtri_2metal}, more metal layers are necessary to realize designs with numerous and isolated or ``islanded'' electrodes, such as those in \cite{kielpinski}.

\begin{figure}[h]
 \centering
 \includegraphics[scale = .50]{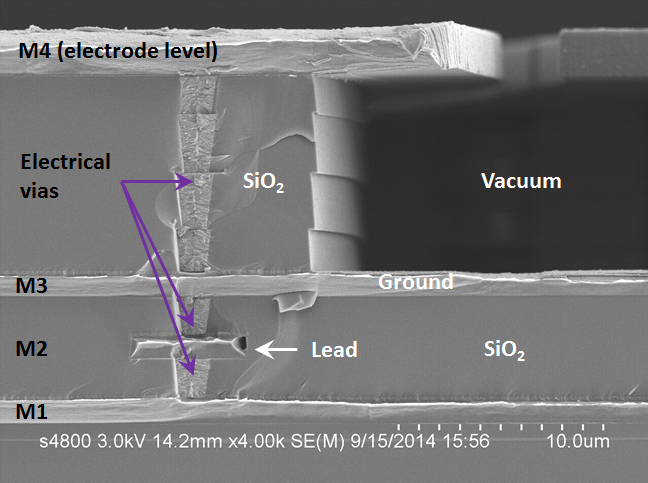}
 % File02.png: 0x0 pixel, 0dpi, 0.00x0.00 cm, bb=
 \caption{Cross-sectional SEM of the trap, showing the four metal levels, interconnects, and insulators.}
 \label{fig:tech}
\end{figure}

To realize access to island electrodes, the preferred number of metal layers is four. We fabricate the four metal level surface trap by using aluminum(Al)-1/2\% copper (Cu) metallization, planarized silicon dioxide (SiO$_2$) inter-level dielectric (ILD), and vertical electrical connections using tungsten plug technology. The layers of metal and ILD are planarized so that all electrode surfaces are flat and reside in the same two dimensional plane. The thin films are integrated on the 25 \micron top silicon (Si) layer of a silicon-on-insulator (SOI) substrate having a 600 \micron Si handle and 2 \micron buried oxide. The first metal level (M1) contains I/O bond pads as well as an rf ground plane to electrically shield the rf leads of the trap structure from the bulk (lossy) Si substrate structure. Metal levels two (M2) and three (M3) allow for lead cross-overs and are used for routing wires from the I/O bond pads to the trap electrodes. Metal level four (M4) covers the entire trap chip area, save for the gaps between rf leads, control electrodes, grounded regions, and at the I/O bond pads in M1 at the edges of the chip. In the trapping region the gaps in M4 reveal the ground layer in M3 which screens the trapped ions from the underlying control electrode potentials on nearby leads in M2. Grounded M4 away from the trap region covers the routing wires in M2 and M3.  Silicon dioxide (SiO$_2$) insulates adjacent metal levels from each other and serves as a mechanical support for electrodes. Using a MEMS "release" etch process, the SiO$_2$ is controllably removed at the gap between trap electrodes in the trapping region as well as at the loading hole, such that the trap electrodes overhang the SiO$_2$ (Fig. \ref{fig:tech}).

The absence of dielectric laterally between electrodes serves to screen the ion from any adventitious charge on the underlying dielectric and allows for evaporation of a different (arbitrary) metal on the top electrodes without shorting. The release etch process removes sacrificial SiO$_2$ which simultaneously singulates the trap chip from the Si substrate wafer and opens a hole through the chip that allows for backside loading of ions. 

For the ring trap used in these experiments, the top metal layer is separated from the bottom metal layer by 14 \micron of ILD and 2.4 \micron (nominal) of metal levels M2 and M3. The separation between M4 and M3 is 10 \micron. M1 and M4 are nominally 2.4 \micron thick while M2 and M3 are nominally 1.2 \micron thick. The two rf rails have a width of 60 \micron and are separated by 134 \micron. The total rf capacitance to ground is $\sim$7 pF. The trap investigated has an additional 50 nm gold (Au) layer (with underlying titanium/platinum adhesion/barrier layer) evaporated onto the Al-1/2\% Cu electrode surface. The backside of the trap is evaporated with 500 nm of Au at an angle to coat the Si side walls of the loading hole, thus pinning the potential on this surface to ground. The trap chip is mounted on top of a gold coated 1.5 mm thick alumina spacer attached to the base of the CPGA package such that the surface of the trap chip is 1.2 mm above that of the package. A conductive cyanate ester adhesive \footnote{Johnson Matthey Electronics JM7000} is used to attach these elements. I/O wirebonds originate from bondpads in M1 and feature low profile wirebonds, protruding a maximum of 30 \micron above the trap electrode plane. I/O bondpads are placed at the perimeter of the trap chip and wire loop heights are minimized to reduce scattering of the lasers from the wirebonds and allow unobstructed optical access across the chip surface. %On the package for one of the gold-coated traps, a filter capacitor of $\sim$1 nF was installed between each control electrode and ground to suppress the pick-up of residual rf voltages, which is crucial for eliminating ion micromotion.

%\section{Experimental setup}
The chip package is mounted into a zero insertion force socket to an in-vacuum filter board with one single pole low pass filter per control electrode. The 89 control voltages are generated outside the vacuum chamber with digital to analog conversion cards and are subject to active and passive filtering before delivery to a chamber feed-through. The rf voltage is generated by a direct digital synthesizer, amplified, and delivered to a chamber feed-through after a step-up by a helical quarter-wave resonator just outside the vacuum vessel. The objective imaging lens is mounted on a translation stage equipped with motorized screws. In the plane of the trap, a circular profile cooling beam (397 nm) is also mounted on a motorized screw. A second cooling beam, passed through a cylindrical lens to yield an elliptical profile, is rotated at about 90$^\circ$ from the first one in the plane of the trap. The circular cooling beam is used for all measurements, while the elliptical cooling beam is used to cover the entire 1.3 mm wide trapping volume along with re-pump beams at 866 nm and at 854 nm, oriented at 45$^\circ$  in the plane of the trap and passed through a cylindrical lens. A 729 nm beam is approximately co-linear with the circular cooling beam and is used for sideband cooling. Photo-ionization of Ca is achieved with 423 nm and 375 nm light. All lasers, motors, rf and control voltages, and imaging components are adjusted and synchronized through Python-based control software.

%\section{Discussion}
In regard to stray field sources, in many instances we are able to roughly correlate field direction with the presence of particulate contaminants on the trap surface. We attribute this to particle charging due to the short wavelength lasers \cite{harlander}. Particle locations are confirmed for select locations by scattering light off the trap surface and independently by scanning electron microscopy (SEM) imaging. One important question about the feasibility of the field correction procedure is the persistence of stray fields in time. Over the course of several months, we found that the stray field does not vary substantially at the measured locations, with the exception of near the loading hole (at $g43$, $g00$, and $g01$). This is not surprising as that location is more exposed to Ca photo-ionization UV light. Given that the automated measurement procedure takes hours, we conclude that the field correction procedure is a reasonable method for engineering an arbitrary field, in our case a corrective one, over a large trapping volume. The results may be improved by exploring different weighting strategies to benefit from the trap layout, or iterative methods to offset trap simulation deficiencies.

\section{Conclusion}
In this work a ring shaped surface trap is demonstrated and characterized by storing about 400 ions with near-uniform spacing over $\sim$90\% of the ring. Equidistant spacing is achieved by measuring and reducing the axial electric fields at a 90 \micron period around the ring, setting a long-range length scale (exceeding the 9 \micron ion spacing) for the background electric fields.  Achieving this uniformity is important for experiments requiring long chains of ions with circular boundary conditions, and could be improved by eliminating the macroscopic surface contaminants and further reducing the size of the loading hole. The presence of particulate surface contaminants is noted by laser scatter in the device under vacuum and subsequently by SEM. These could be eliminated by limiting handling of the device and chamber to clean rooms and by employing a less turbulent pump-down procedure. The issue of the loading hole could be addressed either by shrinking the size of the hole or eliminating it entirely and loading ions between the gap in a split-center electrode. 

\section{Acknowledgments}
BT thanks Hartmut H\"{a}ffner for ultra high vacuum advice and Kevin Fortier for help with the lasers.

Sandia National Laboratories is a multi-program laboratory managed and operated by Sandia Corporation, a wholly owned subsidiary of Lockheed Martin Corporation, for the US Department of Energy's National Nuclear Security Administration under contract DE-AC04-94AL85000
This work is part of the Multi-Qubit Coherent Operations (MQCO) program supported by the Intelligence Advanced Research Projects Activity (IARPA). 

\bibliography{bibliography}
\end{document}